%% file: ServiceChain.tex
\documentclass[conference]{IEEEtran}
\IEEEoverridecommandlockouts

\input{macros}
    
\begin{document}

\title{Delay-Optimal Service Chain Forwarding and Offloading in Collaborative Edge Computing
}

\author{\IEEEauthorblockN{Jinkun Zhang \,and\, Edmund Yeh}
\IEEEauthorblockA{Department of Electrical and Computer Engineering\\
Northeastern University, US}
}

\maketitle

\begin{abstract}
Collaborative edge computing (CEC) is an emerging paradigm for heterogeneous devices to collaborate on edge computation jobs.
For congestible links and computing units, delay-optimal forwarding and offloading for service chain tasks (e.g., DNN with vertical split) in CEC remains an open problem.
In this paper, we formulate the service chain forwarding and offloading problem in CEC with arbitrary topology and heterogeneous transmission/computation capability, and aim to minimize the aggregated network cost. 
We consider congestion-aware nonlinear cost functions that cover various performance metrics and constraints, such as average queueing delay with limited processor capacity.
We solve the non-convex optimization problem globally by analyzing the KKT condition and proposing a sufficient condition for optimality.
We then propose a distributed algorithm that converges to the global optimum. The algorithm adapts to changes in input rates and network topology, and can be implemented as an online algorithm.
Numerical evaluation shows that our method significantly outperforms baselines in multiple network instances, especially in congested scenarios.
\end{abstract}

\section{Introduction}
Recent years have seen an explosion in the number of mobile and IoT devices. Many of the emerging mobile applications, such as VR/AR, autonomous driving, are computation-intensive and time-critical.
Directing all computation requests and their data to the central cloud is becoming impractical due to limited backhaul bandwidth and high associated latency.
Edge computing has been proposed as a promising solution to provide computation resources and cloud-like services in close proximity to mobile devices.   
An extension of the idea of edge computing is the concept of collaborative edge computing (CEC).
In addition to point-to-point offloading, CEC permits multiple stakeholders (mobile devices, IoT devices, edge servers, and cloud) to collaborate with each other by sharing data, communication resources, and computation resources to accomplish computation tasks \cite{sahni2020multi}. 
Devices equipped with computation capabilities can collaborate with each other through D2D communication \cite{sahni2017edge}. Edge servers can also collaborate with each other for load balancing or further with the central cloud to offload demands that they cannot accommodate \cite{zhu2017socially}. 
CEC is also needed if no direct connection exists between devices and edge servers. That is, computation-intensive tasks of unmanned aerial vehicle (UAV) swarms or autonomous cars far from the wireless access point should be collaboratively computed or offloaded through multi-hop routing to the server with the help of other devices \cite{hong2019multi,sahni2017edge}. 

\begin{figure}[htbp]
\centerline{\includegraphics[width=0.4\textwidth]{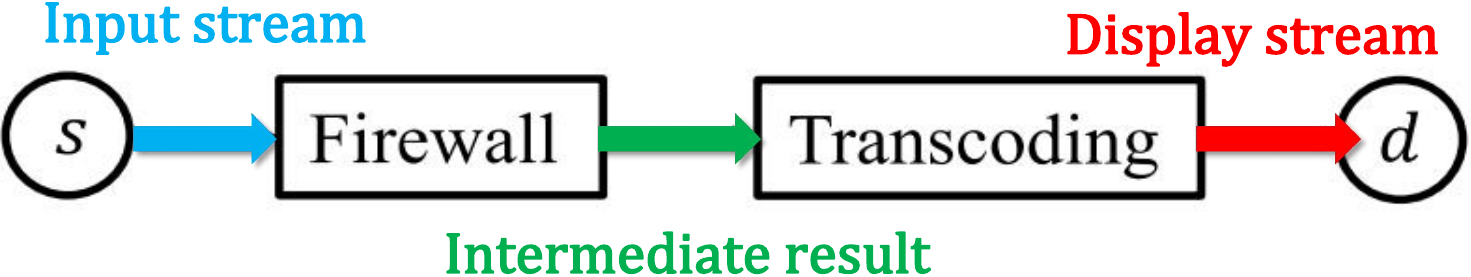}}
\caption{Service chain example: LAN video stream client for from source $s$ to display $d$.  Input data stream undergoes sequential tasks.}
\label{fig_service_chain}
\end{figure}

The delay-optimal routing and offloading for independent one-step computation tasks in CEC has been proposed \cite{zhang2022optimal}.
As a generalization of one-step computation, \emph{service chaining} enables the ordering of computation tasks\cite{han2015network}.
In the service chain model, the network provides services called \emph{applications}, where one application consists of a chain of \emph{tasks}
that are performed sequentially, mapping input data to output results via (potentially multiple) intermediate stages. 
Service chaining has been widely studied in the network function virtualization (NFV) context, e.g., \cite{zhang2021optimal}.
An example of service chaining is shown in Fig \ref{fig_service_chain}. 

Optimization for service chain computation allocation has been widely considered.
Zhang et al. \cite{zhang2021optimal} proposed a throughput-optimal control policy for distributed computing networks with service chain applications and mixed-cast traffic flows, but did not consider the optimization of delay performance.
Khoramnejad et al. \cite{khoramnejad2021distributed} minimized the cost of service chain delay and energy consumption at the edge via deep learning, but within the restricted setting of bipartite network topologies.
Sahni et al. \cite{sahni2018data} optimized for the earliest finish time (EFT) of service chains in mesh networks, albeit without considering link congestion.
Recently, an efficient framework for dispersed computing of interdependent tasks was established by Ghosh et al. \cite{ghosh2021jupiter}, but without a theoretical performance guarantee.

In CEC with arbitrary topology and heterogeneous nodes, the delay-optimal forwarding and offloading of service chain tasks remains an open problem. One major challenge is incorporating congestion-dependent non-linear costs (e.g., queueing delay at transmission links and computing units).
Furthermore, the forwarding and offloading mechanism should preferably be distributed for network scaling.

To meet these challenges, this paper considers an arbitrary multi-hop network, where nodes collaboratively finish multiple service chain applications. 
Nodes have heterogeneous computation capabilities, while some are also data sources (e.g., sensors and mobile users).
For each application, data enter the network at the source nodes. 
Intermediate results are produced and forwarded at collaborative nodes, and the final result is delivered to the destination node. 
To more accurately model the delay incurred on links and computing units, we consider general congestion-dependent non-linear costs. 
We propose a framework that unifies the mathematical representation of forwarding and computation offloading, and tackle the non-linear cost minimization problem with a node-based perspective first introduced by \cite{gallager1977minimum} and adopted in \cite{zhang2022optimal}.
We first characterize the Karush–Kuhn–Tucker (KKT) necessary conditions for the proposed optimization problem, and demonstrate that the KKT condition can lead to arbitrarily suboptimal performance in certain degenerate cases.
To overcome this issue, we propose a modification to the KKT condition, and prove that the new condition is a sufficient for global optimality.
We also devise a distributed and adaptive algorithm that converges to an operating point satisfying the sufficiency condition.

We summarize our detailed contributions as follows:
\begin{itemize}
    \item We investigate joint forwarding and offloading for service chain applications in arbitrary multi-hop network topologies with non-linear transmission and computation costs and formulate a non-convex cost minimization problem.
    \item We study the KKT necessary condition for optimality in the non-convex problem, and present a sufficient condition for optimality by modifying the KKT condition. 
    \item We devise a distributed and adaptive algorithm that converges to the global optimum.
    \item Through numerical evaluation, we demonstrate the advantages of our algorithm relative to baselines in different network instances, especially in congested scenarios.
\end{itemize}
In Section \ref{Section:model} we present the network model and the optimization problem. In Section \ref{Section:conditions} we present the sufficient optimality conditions. In Section \ref{Section:algorithm} we develop a distributed algorithm, and in Section \ref{Section:simulation} we present our numerical results.

\begin{figure}[t]
\centerline{\includegraphics[width=0.4\textwidth]{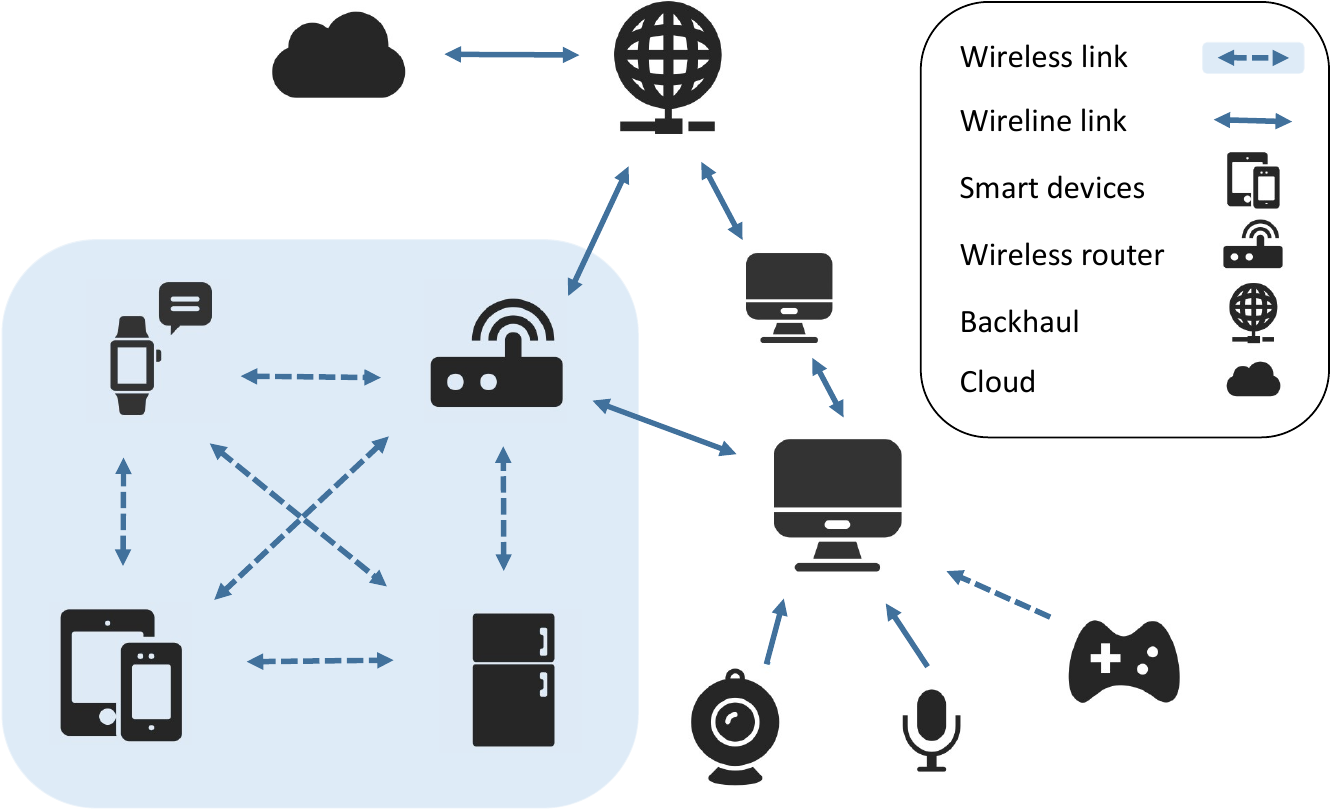}}
\caption{Sample topology involving IoT network on the edge \cite{zhang2022optimal}.}
\label{fig: CEC Network}
\end{figure}

\section{Problem formulation}
\label{Section:model}

We model the CEC network with a directed graph $\mathcal{G} = (\mathcal{V},\mathcal{E})$, where $\mathcal{V}$ and $\mathcal{E}$ are the sets of nodes and links, respectively. Fig. \ref{fig: CEC Network} gives an example CEC network.
A node $i \in \mathcal{V}$ can represent a user device, an edge server, or the cloud.
A link $(i,j) \in \mathcal{E}$ represents an available (wireline or wireless) channel from $i$ to $j$. 
We assume the nodes in $\mathcal{V}$ and links in $\mathcal{E}$ have heterogeneous capabilities for computation and transmission.
The computation and transmission in $\mathcal{G}$ are driven by \emph{applications}, where we let $\mathcal{A}$ denote the set of the applications. 
Each application is a service chain, i.e., an application consists of a finite number of \emph{tasks} performed sequentially with a pre-determined order.
For $a \in \mathcal{A}$, we let $\mathcal{T}_a$ be the (ordered) set of $a$'s tasks. 

\begin{table}[t]
\vspace{0.5\baselineskip}
\footnotesize
\begin{tabular}{l | l }
\hline
$\mathcal{G} = (\mathcal{V},\mathcal{E})$ & Network graph $\mathcal{G}$, set of nodes $\mathcal{V}$ and links $\mathcal{E}$\\
$\mathcal{A}$; $d_a$ & Set of applications; result destination of application $a$\\
$\mathcal{T}_a$ & Service chain tasks of application $a \in \mathcal{A}$ \\
$\mathcal{S}$ & Set of all stages $(a,k)$ where $a \in \mathcal{A}$, $k = 0,\cdots,|\mathcal{T}_a|$\\
$L_{(a,k)}$ & Packet size of stage $(a,k) \in \mathcal{S}$\\
$r_i(a)$ & Exogenous input data rate for application $a$ at node $i$ \\
$t_{i}(a,k)$ & Traffic of stage $(a,k)$ at node $i$ \\
$\phi_{ij}(a,k)$ & Fraction of $t_{i}(a,k)$ forwarded to node $j$ (if $j \neq 0$) \\
$\phi_{i0}(a,k)$ & Fraction of $t_{i}(a,k)$ assigned to CPU at $i$ \\
$f_{ij}(a,k)$ & Rate (packet/sec) of stage $(a,k)$ on link $(i,j)$\\
$g_i(a,k)$ & Rate (packet/sec) of stage $(a,k)$ assigned to CPU at $i$\\
$D_{ij}(F_{ij})$ & Transmission cost (e.g. queueing delay) on $(i,j)$\\
$C_i(G_i)$ & Computation cost (e.g. CPU runtime) at node $i$ \\
\hline
\end{tabular}
\caption{Major notations}
\vspace{-0.5\baselineskip}
\label{table:tab1}
\end{table}

We assume that any node in $\mathcal{V}$ can act as a data source or computation site for any application.
The data input rate for application $a$ at node $i$ is $r_i(a)$ (packet/sec)\footnote{We allow applications to have multiple data sources.}
, where each data packet has size $L_{(a,0)}$ (bit).
Each application has one pre-specified destination $d_a \in \mathcal{V}$, to which the final results of the service chain are delivered. 
For application $a$, the data flows are injected into the network from nodes $i$ with $r_i(a) >0$, forwarded in a hop-by-hop manner to nodes that decide to perform the first task of $a$. After the first task, data flows are converted into flows of intermediate results, and are further forwarded for the next task in the service chain. Eventually, final results are delivered to the destination $d_a$.
We assume the flows of application $a$ are categorized into $|\mathcal{T}_a|+1$ \emph{stages}, where stage $(a,k)$ with $k = 0, 1, \cdots, |\mathcal{T}_a|$ represents the (intermediate) results that have finished the $k$-th computation task of $a$. Particularly, we say the data flows are of stage $(a,0)$, and the final results are of stage $(a,|\mathcal{T}_a|)$. 
We assume packets of stage $(a,k)$ are of size $L_{(a,k)}$ (bit), and denote the set of all stages by $\mathcal{S} = \left\{(a,k)\big|a \in \mathcal{A}, k = 0,1\cdots,\left|\mathcal{T}_a\right|\right\}$.

To model the forwarding and computation offloading behavior in $\mathcal{G}$, we adopt a node-based perspective first introduced in \cite{gallager1977minimum} and followed by \cite{zhang2022optimal}.
We let $t_i(a,k)$ denote the \emph{traffic} of stage $(a,k) \in \mathcal{S}$ at node $i$. Specifically, $t_i(a,k)$ includes both the packet rate of stage $(a,k)$ that is generated at $i$ (this can be the input data rate $r_i(a)$ if $k = 0$, or the newly generated intermediate results at $i$'s computation unit if $k \neq 0$), and the packet rate that is forwarded from other nodes to $i$.
We let $\phi_{ij}(a,k) \in [0,1]$ be the \emph{forwarding/offloading} variable for $i,j \in \mathcal{V}$ and $(a,k) \in \mathcal{S}$.
Specifically, of the traffic $t_i(a,k)$ that arrive at $i$, node $i$ forwards a fraction of $\phi_{ij}(a,k)$ to node $k$.
Moreover, if $k \neq |\mathcal{T}_a|$, node $i$ forwards a fraction $\phi_{i0}(a,k)$ of $t_{i}(a,k)$ to its local CPU to perform computation of the $k+1$-th task.\footnote{For coherence, we let $\phi_{ij}(a,k) \equiv 0$ if $(i,j) \not\in \mathcal{E}$, and $\phi_{i0}(a,|\mathcal{T}_a|) \equiv 0$.
Such fractional forwarding can be achieved via various methods, e.g., random packet dispatching with probability $\phi_{ij}(a,k)$. 
More complex mechanisms also exist to stabilize the actual flow rates, e.g.,\cite{zhang2023congestion}.
} 
We assume for every data or intermediate packet consumed at CPU for computation, one and only one next-stage packet is generated accordingly.
Therefore, for $k = 0$, 
\begin{equation*}
    t_i(a,0) = \sum\nolimits_{j \in \mathcal{V}} t_{j}(a,0) \phi_{ji}(a,0) + r_i(a),
\end{equation*}
 and for $k \neq 0$,
\begin{equation*}
    t_i(a,k) = \sum\nolimits_{j \in \mathcal{V}} t_{j}(a,k) \phi_{ji}(a,k) + t_{i}(a,k-1) \phi_{i0}(a,k-1).
\end{equation*}

Let $\boldsymbol{\phi} = [ \phi_{ij}(a,k)]_{(a,k)\in \mathcal{S}, i \in \mathcal{V}, j \in \{0\} \cup \mathcal{V}}$ be the \emph{global forwarding/offlodaing strategy}, with the following constraint 
\begin{align}
    \sum_{j \in \{0\} \cup \mathcal{V}} \phi_{ij}(a,k) = \begin{cases}
        0, &\quad \text{if } k = \left|\mathcal{T}_a\right| \text{ and } i = d_a,
        \\ 1, &\quad \text{otherwise}.
    \end{cases}\label{FlowConservation_phi}
\end{align}

\begin{figure}[t!]
\centerline{
\includegraphics[width=0.43\textwidth]{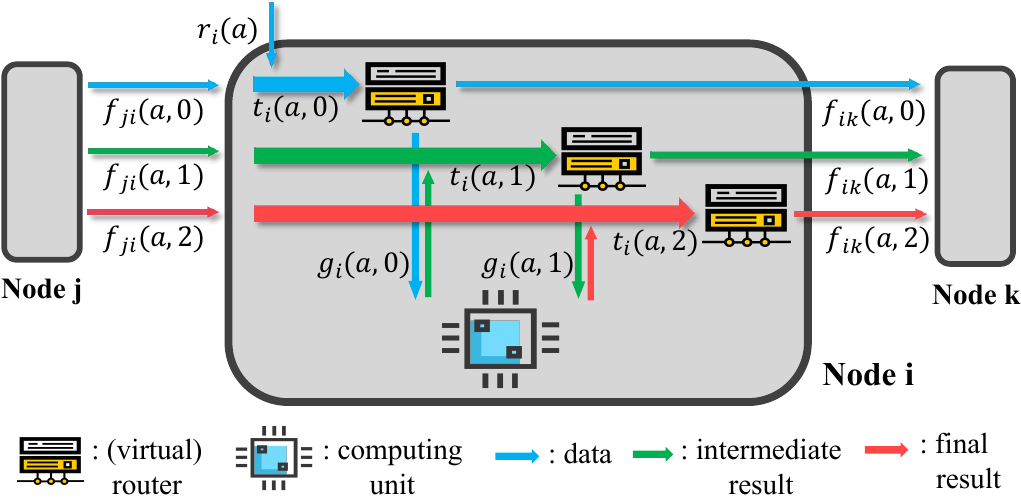}
}
\caption{Detailed behavior of flows on nodes $j,i,k$ when $|\mathcal{T}_a| = 2$. Node $i$ handles traffic $t_i(a,k)$ for different stage $(a,k)$ via (virtual) routers. Each router is controlled by variable $[\phi_{ij}(a,k)]_{j\in\{0\}\cup \mathcal{V}}$.} 
\label{fig_node_behavior}
\end{figure}

Constraint \eqref{FlowConservation_phi} guarantees that all input data will eventually be processed to the final result, and exit the network at the destination. 
Let $f_{ij}(a,k)$ be the packet rate (packet/sec) for stage $(a,k)$ packets transmitted on link $(i,j)$, and let $g_i(a,k)$ be the packet rate (packet/sec) for stage $(a,k)$ packets forwarded to $i$'s CPU for computation. Then,
\begin{equation*}
    f_{ij}(a,k) = t_{i}(a,k) \phi_{ij}(a,k), \quad g_{i}(a,k) = t_{i}(a,k) \phi_{i0}(a,k).
\end{equation*}
The {total flow rate} (bit/sec) on link $(i,j)\in\mathcal{E}$ is given by
\begin{equation*}
    F_{ij} = \sum\nolimits_{(a,k) \in \mathcal{S}} L_{(a,k)} f_{ij}(a,k),
\end{equation*}
and the total computation workload at node $i \in \mathcal{V}$ is given by
\begin{equation*}
    G_i = \sum\nolimits_{ (a,k) \in \mathcal{S}} w_i(a,k) g_i(a,k),
\end{equation*}
where $w_i(a,k)$ is the \emph{computational weight}, i.e., the computation workload for node $i$ to perform the $(k+1)$-th task of application $a$ on a single input packet. 
Fig. \ref{fig_node_behavior} gives a detailed illustration of how the flows are directed within one node. 

Non-linear costs are incurred on the links due to transmission, and on the nodes due to computation.
We denote the transmission cost on link $(i,j)$ by $D_{ij}(F_{ij})$, where function $D_{ij}(\cdot)$ is continuously differentiable, monotonically increasing and convex, with $D_{ij}(0) = 0$.
Such $D_{ij}(\cdot)$ subsumes various existing cost functions, including 
linear cost (transmission delay), link capacity constraints.
It also incorporates congestion-dependent performance metrics.
For example, let $\mu_{ij}$ be the service rate of an M/M/1 queue, then $D_{ij}(F_{ij}) = {F_{ij}}/\left({\mu_{ij}-F_{ij}}\right)$ gives the average number of packets waiting in the queue or being served on $(i,j)$ \cite{bertsekas2021data}. 
The computation cost at $i$ is denoted by $C_i(G_i)$, where $C_i(\cdot)$ is also increasing, continuously differentiable and convex, with $C_i(0) = 0$.
Function $C_i(G_i)$ can incorporate computation congestion (e.g., average number of packets waiting for available processor or being served at CPU).
When both $D_{ij}(F_{ij})$ and $C_i(G_i)$ represent queue lengths, by Little's Law, the network aggregated cost is proportional to the expected packet system delay.
Our major notation is summarized in Table \ref{table:tab1}.

We aim to minimize the aggregated transmission and computation cost in the network, formally cast as\footnote{We do not explicitly impose any link or computation capacity constraints in \eqref{ServiceChainObj}, as they are already incorporated in the cost functions.}
\begin{equation}
\begin{aligned}
    &\min_{\boldsymbol{\phi}} \quad D(\boldsymbol{\phi}) = \sum\nolimits_{(i,j)\in \mathcal{E}} D_{ij}(F_{ij}) + \sum\nolimits_{i \in \mathcal{V}} C_{i}(G_i) 
    \\ &\text{subject to} \quad \phi_{ij}(a,k) \in [0,1], \text{ and \eqref{FlowConservation_phi} holds.}
\end{aligned}
\label{ServiceChainObj}
\end{equation}

\section{Optimality Conditions}
\label{Section:conditions}
To tackle the non-convex problem \eqref{ServiceChainObj}, we first present the KKT necessary condition, and demonstrate that the KKT condition can yield arbitrarily worse performance when compared to that of the global optimum.
Then, we propose a sufficient condition for optimality by modifying the KKT condition. 

We start by giving closed-form derivatives of $D(\boldsymbol{\phi})$.
Our analysis generalizes \cite{zhang2022optimal} and makes non-trivial extensions incorporating service chain flows.
For link $(i,j)$ and stage $(a,k)$, the marginal cost of increasing $\phi_{ij}(a,k)$ is equivalent to a sum of two parts, 1) the marginal communication cost on link $(i,j)$ due to increasing of $f_{ij}(a,k)$, and 2) the marginal cost due to increasing of node $j$'s traffic $t_j(a,k)$.
Formally, for $j \neq 0$,
\begin{subequations}\label{partial_D_phi}
\begin{equation}
    \frac{\partial D}{\partial \phi_{ij}(a,k)} =  t_i(a,k) \left(L_{(a,k)} D^\prime_{ij}(F_{ij}) + \frac{\partial D}{\partial t_j(a,k)}\right),
\label{partial_D_phi_j}
\end{equation}
For $j = 0$, the marginal cost of increasing $\phi_{i0}(a,k)$ consists of 1) marginal computation cost at node $i$, and 2) marginal cost for increasing next-stage traffic $t_i(a,k+1)$. Namely\footnote{For coherence, we let $\partial D/\partial \phi_{ij}(a,k) \equiv \infty$ for $(i,j) \not \in \mathcal{E}$, and $\partial D/\partial \phi_{i0}(a,|\mathcal{T}_a|) \equiv \infty$ as the final results will not be further computed.},
\begin{equation}
    \frac{\partial D}{\partial \phi_{i0}(a,k)} =  t_i(a,k)\left(w_i(a,k) C^\prime_{i}(G_{i}) + \frac{\partial D}{\partial t_i(a,k+1)}\right). 
\label{partial_D_phi_0}
\end{equation}
\end{subequations}

In \eqref{partial_D_phi}, $\partial D/\partial t_i(a,k)$ is a weighted sum of marginal costs for out-going links and local CPU, i.e., if $k \neq |\mathcal{T}_a|$,
\begin{subequations}
\label{partial_D_r}
\begin{align}
    &\frac{\partial D}{\partial t_i(a,k)} = \phi_{i0}(a,k)\left(w_i(a,k) C^\prime_{i}(G_{i}) + \frac{\partial D}{\partial t_i(a,k+1)}\right)\nonumber
   \\ &+ \sum_{j \in \mathcal{V} } \phi_{ij}(a,k)\left(L_{(a,k)} D^\prime_{ij}(F_{ij}) + \frac{\partial D}{\partial t_j(a,k)}\right),\label{partial_D_r_1}
\end{align}
and for $k = |\mathcal{T}_a|$, recall $\phi_{i0}(a,|\mathcal{T}_a|) \equiv 0$, $\frac{\partial D}{\partial t_i(a,|\mathcal{T}_a|)}$ equals
\begin{equation}
\label{partial_D_r_2}
    \sum_{j \in \mathcal{V} } \phi_{ij}(a,|\mathcal{T}_a|)\left(L_{(a,|\mathcal{T}_a|)} D^\prime_{ij}(F_{ij}) + \frac{\partial D}{\partial t_j(a,|\mathcal{T}_a|)}\right).
\end{equation}
\end{subequations}

Recall the fact that final results do not introduce any cost at destination, i.e., $\partial D/\partial t_{d_a}(a,|\mathcal{T}_a|) \equiv 0$, one could calculate $\partial T/\partial t_i(a,k)$ recursively by \eqref{partial_D_r}.
With the absence of computation offloading, a rigorous proof of \eqref{partial_D_phi}, \eqref{partial_D_r} is provided in \cite{gallager1977minimum} Theorem 2.
Then, Lemma \ref{Lemma_Necessary} gives the KKT necessary optimality condition for problem \eqref{ServiceChainObj}.
\begin{lem} [KKT necessary condition]
\label{Lemma_Necessary}
Let $\boldsymbol{\phi}$ be an optimal solution to \eqref{ServiceChainObj}, then for all $i \in \mathcal{V}$, $j\in\left\{0\right\} \cup \mathcal{V}$, and $(a,k)\in \mathcal{S}$,
\begin{equation}
    \frac{\partial D}{ \partial \phi_{ij}(a,k)}  
    \begin{cases}
     = \min\limits_{j^\prime \in \left\{0\right\} \cup \mathcal{V} } \frac{\partial D}{ \partial \phi_{ij^\prime}(a,k)} , \, \text{if } \phi_{ij}(a,k) >0,
    \\ \geq \min\limits_{j^\prime \in \left\{0\right\} \cup \mathcal{V} } \frac{\partial D}{ \partial \phi_{ij^\prime}(a,k)} ,  \, \text{if } \phi_{ij}(a,k) =0.
    \end{cases}
\label{Condition_KKT}
\end{equation}
\end{lem}

\begin{figure}[htbp]
\centerline{\includegraphics[width=0.4\textwidth]{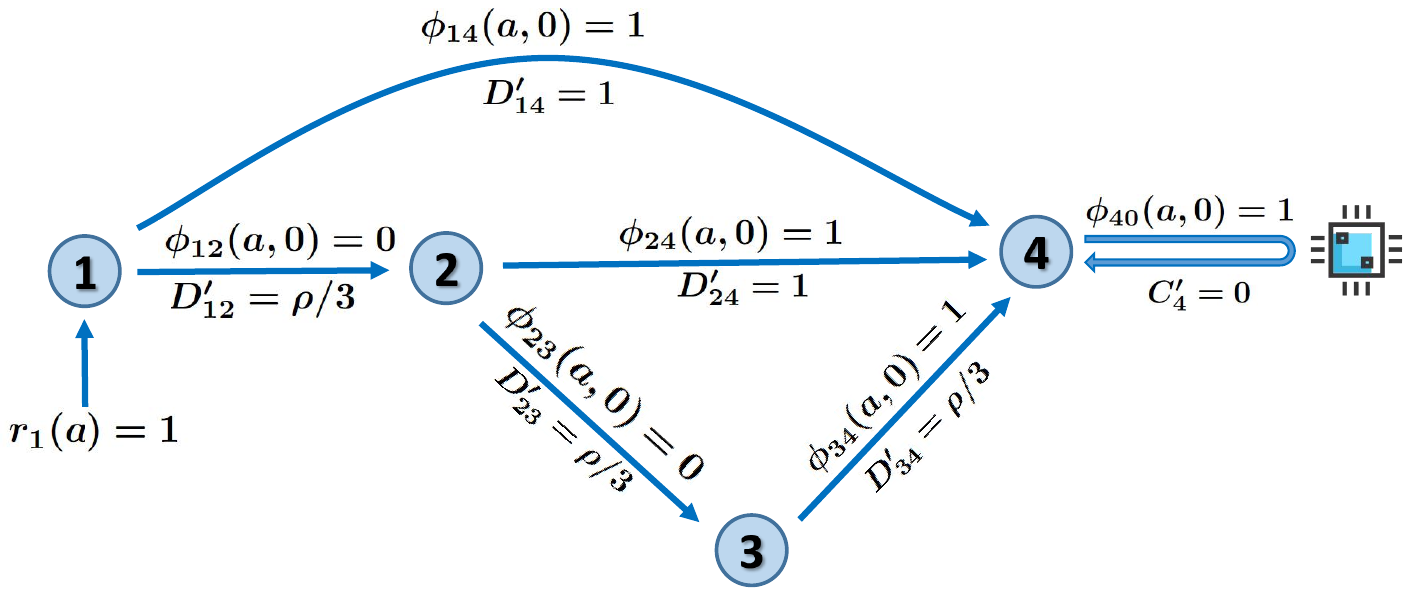}}
\caption{The KKT condition \eqref{Condition_KKT} leads to arbitrarily suboptimal solution. Consider application $a$ with $|\mathcal{T}_a| = 1, d_a = 4$. Data is input at $1$ and CPU is only equipped at $4$. All cost functions are linear. Variable $\boldsymbol{\phi}$ shown in the figure satisfies \eqref{Condition_KKT} with $D(\boldsymbol{\phi}) = 1$. However, the optimal strategy is to forward all data on path $1 \to 2 \to 3 \to 4$, with $D(\boldsymbol{\phi}^*) = \rho$. Thus Proposition \ref{prop_arbitrarily_worse} holds by letting $\rho$ arbitrarily small.
}
\label{fig_kkt_suboptimal}
\end{figure}

Proof of Lemma \ref{Lemma_Necessary} is omitted due to space limit.
Condition \eqref{Condition_KKT} only gives the necessary condition for optimality. In fact, variable $\boldsymbol{\phi}$ satisfying condition \eqref{Condition_KKT} may perform arbitrarily worse than the global optimal solution.
{
\begin{prop}
\label{prop_arbitrarily_worse}
    For any $0<\rho<1$, there exists a scenario (i.e., network $\mathcal{G}$, applications $\mathcal{A}$, cost functions $F_{ij}(\cdot)$, $G_i(\cdot)$, and input rates $r_i(a)$) such that $\frac{D(\boldsymbol{\phi}^*)}{D(\boldsymbol{\phi})} = \rho$, where $\boldsymbol{\phi}$ is feasible to \eqref{ServiceChainObj} and satisfies \eqref{Condition_KKT}, and $\boldsymbol{\phi}^*$ is an optimal solution to \eqref{ServiceChainObj}.
\end{prop}

A scenario with $\frac{D(\boldsymbol{\phi}^*)}{D(\boldsymbol{\phi})} = \rho$ is demonstrated in Fig. \ref{fig_kkt_suboptimal}.
}
The suboptimality of condition \eqref{Condition_KKT} is incurred by degenerate cases at $i$ and $(a,k)$ satisfying $t_i(a,k) = 0$, where \eqref{Condition_KKT} automatically holds regardless of the actual forwarding variables $\phi_{ij}(a,k)$.

We now address the non-sufficiency of condition \eqref{Condition_KKT}.
Inspired by \cite{zhang2022optimal}, we introduce a modification to \eqref{Condition_KKT} that leads to a sufficient condition for optimality in \eqref{ServiceChainObj}.
Observing that for any $i$ and $(a,k)$, the traffic term $t_i(a,k)$ repeatedly exists in RHS of \eqref{partial_D_phi} for all $j \in \{0\} \cup \mathcal{V}$. Thus when $t_i(a,k) = 0$, it always holds that $\phi_{ij}(a,k) = 0$. 
We thereby remove the traffic terms in \eqref{Condition_KKT}, leading to condition \eqref{Condition_sufficient}. 

\begin{theo}[Sufficiency condition]
\label{Thm_Sufficient}
Let $\boldsymbol{\phi}$ be feasible to \eqref{ServiceChainObj}. Then $\boldsymbol{\phi}$ is a global optimal solution to \eqref{ServiceChainObj} if the following holds for all $i \in \mathcal{V}$, $j\in\left\{0\right\} \cup \mathcal{V}$ and $(a,k) \in \mathcal{S}$,
\begin{equation}
    \delta_{ij}(a,k) \begin{cases}
    = \min\limits_{j^\prime \in \left\{0\right\}  \cup \mathcal{V}} \delta_{ij^\prime}(a,k), \, \text{if } \phi_{ij}(a,k) >0,
    \\ \geq \min\limits_{j^\prime \in \left\{0\right\}  \cup \mathcal{V}} \delta_{ij^\prime}(a,k), \, \text{if } \phi_{ij}(a,k) =0,
    \end{cases}
\label{Condition_sufficient}
\end{equation}

where $\delta_{ij}(a,k)$ is the ``modified marginal", given by
\begin{equation}
    \delta_{ij}(a,k) = \begin{cases}
        L_{(a,k)} D^\prime_{ij}(F_{ij}) + \frac{\partial D}{\partial t_j(a,k)}, \quad &\text{if } j \neq 0,
        \\ w_i(a,k) C^\prime_{i}(G_{i}) + \frac{\partial D}{\partial t_i(a,k+1)}, \quad &\text{if } j = 0.
    \end{cases}
\label{delta}
\end{equation}
\end{theo}
Proof of Theorem \ref{Thm_Sufficient} is provided in our supplementary material \cite{zhang2023delayoptimal}. 
To see the difference between the KKT necessary condition \eqref{Condition_KKT} and the sufficiency condition \eqref{Condition_sufficient}, consider again the example in Fig. \ref{fig_kkt_suboptimal}.
For any $\boldsymbol{\phi}$ satisfying \eqref{Condition_sufficient}, it must hold that $\phi_{12}(a,0) = 1$, $\phi_{23}(a,0) = 1$ and $\phi_{34}(a,0) = 1$, precisely indicating the shortest path $1 \to 2 \to 3 \to 4$ as expected.

\section{Distributed Algorithm }
\label{Section:algorithm}
We propose a distributed algorithm that converges to the sufficiency condition \eqref{Condition_sufficient}.
The algorithm generalizes the method in \cite{gallager1977minimum} to service chain computation placement. 
It is adaptive to changes in data input rates and network topology, and can be implemented as an online algorithm.

Existence of routing loops generates redundant flow circulation and causes potential instability. 
We say $\boldsymbol{\phi}$ has a \emph{loop} of stage $(a,k)$ if there exists $i$, $j \in \mathcal{V}$, such that $i$ has a path\footnote{A \emph{path} from $i$ to $j$ is a sequence of nodes $n_1, \cdots ,n_L$ with $(n_l,n_{l+1}) \in \mathcal{E}$ and $\phi_{n_l n_{l+1}}(a,k) > 0$ for $l = 1,\cdots,L-1$, and $n_1 = i$, $n_L = j$.} of stage $(a,k)$ to $j$, and vice versa.
We say $\boldsymbol{\phi}$ is \emph{loop-free} if no loops are formed for any stage $(a,k) \in \mathcal{S}$. 
We assume the network starts with a feasible and loop-free strategy ${\boldsymbol{\phi}}^0$, and the initial cost $D^0 = D({\boldsymbol{\phi}}^0)$ is finite.
Time is partitioned into slots of duration $T$. At $(t+1)$-th slot $(t \geq 0)$, node $i$'s forwarding/offloading strategy $\boldsymbol{\phi}_{i}$ is updated as follows,
\begin{equation}
     \boldsymbol{\phi}_i^{t+1} = \boldsymbol{\phi}_i^{t} + \Delta\boldsymbol{\phi}_i^{t},
\label{variable_update}
\end{equation}
where the update vector $\Delta\boldsymbol{\phi}_i^{t}$ is calculated by
\begin{equation}
\begin{aligned}
    &\Delta\phi_{ij}^{t}(a,k) =
\\    &\begin{cases}
    - \phi_{ij}^{t}(a,k), & \text{if } j \in \mathcal{B}_i^t(a,k)
    \\ S_i^t(a,k) / N_i^t(a,k),  & \text{else if } e_{ij}^t(a,k) = 0
    \\ -\min\left\{ \phi_{ij}^t(a,k), \alpha e_{ij}^t(a,k) \right\}, & \text{else if } e_{ij}^t(a,k) > 0
    \end{cases}
\end{aligned}
\label{dphi_and_dy}
\end{equation}
where $\mathcal{B}_i^t(a,k)$ is the set of \emph{blocked nodes} to suppress routing loops
, $\alpha$ is the stepsize, and
\begin{align}
     & e_{ij}^t(a,k) = \delta_{ij}^t(a,k) - \min\nolimits_{j \not\in \mathcal{B}_i^t(a, k)} \delta_{ij}^t(a,k), \, \forall j \not\in \mathcal{B}_i^t(a,k), \nonumber
     \\ & N_i^t(a,k) = \bigg|\left\{j \not\in \mathcal{B}_i^t(a, k) \big| e_{ij}^t(a,k) = 0 \right\}\bigg|, \label{algorithm_detail_eNS}
     \\ & S_i^t(a, k)= \sum\nolimits_{j: e_{ij}^t(a,k) > 0}\Delta\phi_{ij}^{t}(a, k). \nonumber
\end{align}

Our method can be viewed as a variant of gradient projection.
It transfers $\phi_{ij}(a,k)$ from non-minimum-marginal directions to the minimum-marginal ones  (here ``marginal" refers to $\delta_{ij}(a,k)$ defined in \eqref{delta}), until \eqref{Condition_sufficient} is satisfied.
We next provide details regarding the blocked node set $\mathcal{B}_{i}^t(a,k)$, and introduce a distributed online mechanism to estimate $\delta_{ij}^t(a,k)$.

\vspace{0.5\baselineskip}
\noindent\textbf{Blocked node set.} To ensure feasibility and loop-free, we adopt the method of blocked node sets following \cite{gallager1977minimum}.
Combining Theorem \ref{Thm_Sufficient} with \eqref{partial_D_r}, if $\boldsymbol{\phi}$ is a global optimal solution satisfying \eqref{Condition_sufficient}, for any stage $(a,k)$, the value of $\partial D/\partial t_i(a,k)$ should decrease monotonically along any path of stage $(a,k)$. Thus node $i$ should not forward flow of stage $(a,k)$ to a neighbor $j$ if either 1) $\partial D/\partial t_j(a,k) > \partial D/\partial t_i(a,k)$, or 2) $j$ could form a path containing some link $(p,q)$ such that $\partial D/\partial t_q(a,k) > \partial D/\partial t_p(a,k)$. 
The set containing nodes of these two categories, along with $j$ with $(i,j) \not\in \mathcal{E}$, is marked as the \emph{blocked node set} $\mathcal{B}_i(a,k)^t$.
Practically, the information needed to determine blocked node sets could be piggy-backed on the broadcast messages described shortly. If such a blocking mechanism is implemented in each iteration, the loop-free property is guaranteed to hold throughout the algorithm. 

\vspace{0.5\baselineskip}
\noindent\textbf{Marginal cost broadcast.}
Recall from \eqref{delta} that in order to calculate $\boldsymbol{\delta}_{i}(a,k)$, node $i$ needs the link marginal costs $D^\prime_{ij}(F_{ij})$, computation marginal costs $C^\prime_i(G_i)$, and the marginal costs due to traffic term, i.e., $\partial D/ \partial t_j(a,k)$ and $\partial D/ \partial t_i(a,k+1)$.
Suppose $D_{ij}(\cdot)$ and $C_i(\cdot)$ are known in closed-form, nodes can directly measure $D^\prime_{ij}(F_{ij})$ and $C^\prime_{i}(G_i)$ while transmitting on link $(i,j)$ and performing computation (or equivalently, first measure flows $F_{ij}$ and workloads $G_i$, then substitute into $D_{ij}(\cdot)$ and $C_i(\cdot)$). 
To collect $\partial D/ \partial t_i(a,k)$, we use recursive calculate \eqref{partial_D_r} starting with $i = d_a$ and $k = |\mathcal{T}_a|$ satisfying $\partial D / \partial t_{d_a}(a,|\mathcal{T}_a|) = 0$.
To carry out this recursive calculation, we apply a multi-stage distributed broadcast protocol to every application $a \in \mathcal{A}$, described as follows:

\begin{enumerate}
    \item Broadcast of $\partial D/\partial t_i(a,|\mathcal{T}_a|)$: 
    Each node $i$ first waits until it receives messages carrying $\partial D/\partial t_j(a,|\mathcal{T}_a|)$ from all its downstream neighbors $j\in\mathcal{V}$ with $\phi_{ij}(a,|\mathcal{T}_a|) > 0$. 
    Then, node $i$ calculates $\partial D/\partial t_i(a,|\mathcal{T}_a|)$ by \eqref{partial_D_r_2} with the measured $D^\prime_{ij}(F_{ij})$ and received $\partial D/\partial t_j(a,|\mathcal{T}_a|)$. 
    Next, node $i$ broadcasts newly calculated $\partial D/\partial t_i(a,|\mathcal{T}_a|)$ to all its upstream neighbors $k\in \mathcal{V}$ with $\phi_{ki}(a,|\mathcal{T}_a|) > 0$. 
    (This stage starts with the destination $d_a$, where $d_a$ broadcasts $\partial D/\partial t_{d_a}(a,|\mathcal{T}_a|) = 0$ to its upstream neighbors.) 
     
    \item Broadcast of $\partial D/\partial t_i(a,k)$ for $k \neq |\mathcal{T}_a|$:
    (This stage starts with $k = |\mathcal{T}_a|-1$ and is repeated with $k$ being abstracted by $1$ each time, until $k = 0$.)
    Suppose every node $i$ has calculated $\partial D/\partial t_i(a,k^\prime)$ for all $k^\prime \geq k+1$. 
    Then, similar as stage 1), $\partial D/\partial t_i(a,k)$ can be calculated recursively by \eqref{partial_D_r_1} via broadcast.
    Besides $\partial D/\partial t_j(a,k)$ from all downstream neighbors $j$, node $i$ must also obtain $\partial D/\partial t_i(a,k+1)$ and $C^\prime_i(G_i)$ for \eqref{partial_D_r_1}.
    For each $k$, this stage starts at nodes $i$ where $\phi_{ij}(a,k) = 0$ for all $j \in \mathcal{V}$.{ i.e., the end-nodes of stage $(a,k)$ paths.} 
\end{enumerate}

With loop-free guaranteed, the broadcast procedure above is guaranteed to traverse throughout the network for all $(a,k) \in \mathcal{S}$ 
and terminate within a finite number of steps. 
We summarize our proposed method in Algorithm \ref{alg1}.

\vspace{0.2\baselineskip}
\setlength{\textfloatsep}{0pt}
\begin{algorithm}
\SetKwRepeat{DoDuring}{do}{during}
\SetKwRepeat{DoAt}{do}{at}
\SetKwRepeat{DoWhen}{do}{when}
Start with $t = 0$ and a loop-free $\bf{\phi}^0$ with $D^0 < \infty$.\\
\DoWhen{ \text{end of iteration $t$}}
{
Each node $i$ obtain $\partial D/\partial t_i(a,k)$ for all $(a,k) \in \mathcal{S}$ via the marginal cost broadcast.\\
Node $i$ calculates $\delta_{ij}(a,k)^t$ by \eqref{delta} for $j \in \{0\}\cup\mathcal{V}$.\\
Node $i$ calculates the blocked node sets $\mathcal{B}_{ij}(a,k)^t$. \\
Node $i$ obtain $\Delta\boldsymbol{\phi}_i^t$ by \eqref{dphi_and_dy} and updates $\boldsymbol{\phi}_i^t$ by \eqref{variable_update}.\\
}
\caption{Gradient Projection (GP)}
\label{alg1}
\end{algorithm}

\vspace{-0.5\baselineskip}
\noindent\textbf{Convergence and complexity.}
The proposed algorithm can be implemented in an online fashion as it does not require prior knowledge of data input rates $r_i(a)$.
Moreover, it is adaptive to changes in $r_i(a)$ since flow rates $F_{ij}$ and workload $G_i$ can be directly estimated by the packet number on links and CPU in previous time slots. 
It can also adapt to changes in the network topology: whenever a link $(i,j)$ is removed from $\mathcal{E}$, node $i$ only needs to add $j$ to the blocked node set; when link $(i,j)$ is added to $\mathcal{E}$, node $i$ removes $j$ from the blocked node set.\footnote{When a new node $v$ is added to the network, it can randomly initiate $\boldsymbol{\phi}_v$ with \eqref{FlowConservation_phi}. $t_v(a,k)$ will be automatically updated by the marginal cost broadcast in the next time slot, and the loop-free property will be guaranteed.}

\begin{theo}
\label{thm_convergence}
Assume the network starts at $\boldsymbol{\phi}^0$ with $D^0 < \infty$, and $\boldsymbol{\phi}^t$ is updated by Algorithm \ref{alg1} with a sufficiently small stepsize $\alpha$. 
Then, the sequence $\left\{\boldsymbol{\phi}^t\right\}_{t = 0}^{\infty}$ converges to a limit point $\boldsymbol{\phi}^*$ that satisfies condition \eqref{Condition_sufficient}.
\end{theo}

The proof of Theorem \ref{thm_convergence} is a straightforward extension of \cite{gallager1977minimum} Theorem 5, and is omitted due to space limitation.
The stepsize $\alpha$ that guarantees convergence can also be found in \cite{gallager1977minimum}.
We remark that the convergence property of Algorithm \ref{alg1} can be improved by adopting second-order quasi-Newton methods, e.g., the method in \cite{zhang2022optimal} speeds up convergence while guaranteeing convergence from any initial point. 

Recall that $\boldsymbol{\phi}$ is updated every time slot of duration $T$, every broadcast message is sent once in every slot.
Thus there are $|\mathcal{E}|$ broadcast message transmissions for each stage in one slot, and totally $|\mathcal{S}||\mathcal{E}|$ per slot, with on average $|\mathcal{S}|/T$ per link/second, and at most $|\mathcal{S}|\Bar{d}$ for each node, where $\Bar{d}$ is the largest out-degree.
We assume the broadcast messages are sent in an out-of-band channel.
Let $t_c$ be the maximum transmission time for a broadcast message, and
$\Bar{h}$ be the maximum hop number for all paths. 
The broadcast completion time for application $a$ is at most $(|\mathcal{T}_a|+1)\Bar{h}t_c$.
Moreover, the proposed algorithm has space complexity $O(|\mathcal{S}|)$ at each node. 
The update may fail if broadcast completion time exceeds $T$ or $|\mathcal{S}|/T$ exceeds the broadcast channel capacity. 
If so, we can use longer slots, or allow some nodes to perform updates every multiple slots.
Meanwhile, if $|\mathcal{S}|$ is large, the algorithm overhead can be significantly reduced by applying our algorithm only to the top applications causing most network traffic.

\section{Numerical evaluation}
\label{Section:simulation}

\begin{figure*}
\centerline{\includegraphics[width=0.9\textwidth]{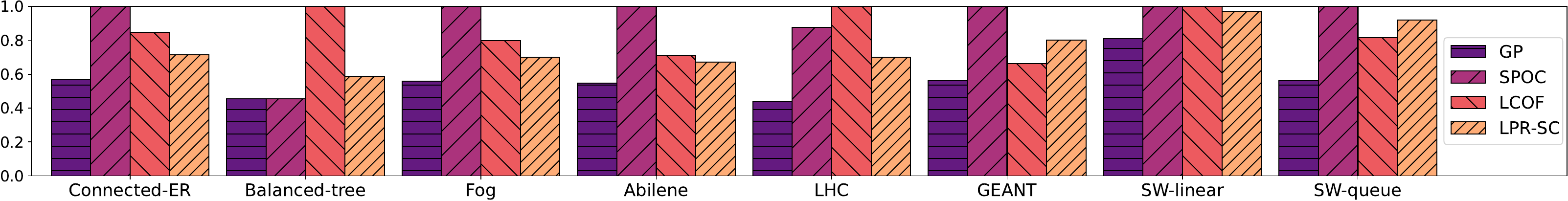}}
\caption{Normalized total cost for network scenarios in Table \ref{tab_scenario}}
\label{fig_bar}
\end{figure*}

We evaluate the proposed algorithm \textbf{GP} via a flow-level simulator presented at \cite{Zhang_Joint-Routing-and-Computation-2022_2022}. 
We implement several baselines and compare against \texttt{GP} over different network scenarios summarized in Table \ref{tab_scenario}. Specifically, \textbf{Connected-ER} is a connectivity-guaranteed Erdős-Rényi graph.
\textbf{Balanced-tree} is a complete binary tree.
\textbf{Fog} is a sample topology for fog-computing \cite{kamran2019deco}.
\textbf{Abilene} is the predecessor of \emph{Internet2 Network}.
\textbf{GEANT} is a pan-European research and education data network.
\textbf{SW} (small-world) is a ring-like graph with additional short-range and long-range edges.
Table \ref{tab_scenario} also summarizes the number of nodes $|\mathcal{V}|$, edges $|\mathcal{E}|$, and stages $|\mathcal{S}|$.
We assume each application has $R$ random active data sources (i.e., the nodes $i$ for which $r_i(a) > 0$), and $r_{i}(a,k)$ for each data source is chosen u.a.r. in $[0.5, 1.5]$.
\textbf{Link} is the type of $D_{ij}(\cdot)$, where \emph{Linear} denotes a linear cost $D_{ij}(F_{ij}) = d_{ij}F_{ij}$, and \emph{Queue} denotes the non-linear cost $D_{ij}(F_{ij}) = \frac{F_{ij}}{d_{ij} - F_{ij}}$.
\textbf{Comp} is the type of $C_i(G_i)$, where \emph{Linear} denotes  $C_i(G_i) = s_i\sum w_{i}(a,k)g_{i}(a,k)$, and \emph{Queue} denotes $C_i(G_i) = \frac{G_i}{s_i - G_i}$.

\begin{table}[htbp]
\footnotesize
\begin{tabular}{|c|p{0.01\textwidth}|p{0.01\textwidth}|p{0.01\textwidth}|p{0.01\textwidth}|c|p{0.01\textwidth}|c|p{0.015\textwidth}|}
\hline
\textbf{Network}&\multicolumn{8}{c|}{\textbf{Parameters}} \\
\textbf{Topology} & \multicolumn{1}{p{0.01\textwidth}}{$|\mathcal{V}|$} & \multicolumn{1}{p{0.01\textwidth}}{$|\mathcal{E}|$} & \multicolumn{1}{p{0.01\textwidth}}{$|\mathcal{A}|$} & \multicolumn{1}{p{0.01\textwidth}}{$R$} & \multicolumn{1}{c}{\textbf{Link}} & \multicolumn{1}{p{0.015\textwidth}}{$\Bar{d}_{ij}$} &\multicolumn{1}{c}{\textbf{Comp}}& $\Bar{s}_i$ \\
\hline
Connected-ER& \multicolumn{1}{p{0.01\textwidth}}{$20$} & \multicolumn{1}{p{0.01\textwidth}}{$40$} & \multicolumn{1}{p{0.01\textwidth}}{$5$} & \multicolumn{1}{p{0.01\textwidth}}{$3$} & \multicolumn{1}{c}{Queue} & \multicolumn{1}{p{0.015\textwidth}}{$10$} &\multicolumn{1}{c}{Queue}& $12$ \\
Balanced-tree & \multicolumn{1}{p{0.01\textwidth}}{$15$} & \multicolumn{1}{p{0.01\textwidth}}{$14$} & \multicolumn{1}{p{0.01\textwidth}}{$5$} & \multicolumn{1}{p{0.01\textwidth}}{$3$} & \multicolumn{1}{c}{Queue} & \multicolumn{1}{p{0.015\textwidth}}{$20$} &\multicolumn{1}{c}{Queue}& $15$ \\
Fog & \multicolumn{1}{p{0.01\textwidth}}{$19$} & \multicolumn{1}{p{0.01\textwidth}}{$30$} & \multicolumn{1}{p{0.01\textwidth}}{$5$} & \multicolumn{1}{p{0.01\textwidth}}{$3$} & \multicolumn{1}{c}{Queue} & \multicolumn{1}{p{0.015\textwidth}}{$20$} &\multicolumn{1}{c}{Queue}& $17$ \\
Abilene & \multicolumn{1}{p{0.01\textwidth}}{$11$} & \multicolumn{1}{p{0.01\textwidth}}{$14$} & \multicolumn{1}{p{0.01\textwidth}}{$3$} & \multicolumn{1}{p{0.01\textwidth}}{$3$} & \multicolumn{1}{c}{Queue} & \multicolumn{1}{p{0.015\textwidth}}{$15$} &\multicolumn{1}{c}{Queue}& $10$ \\
LHC & \multicolumn{1}{p{0.01\textwidth}}{$16$} & \multicolumn{1}{p{0.01\textwidth}}{$31$} & \multicolumn{1}{p{0.01\textwidth}}{$8$} & \multicolumn{1}{p{0.01\textwidth}}{$3$} & \multicolumn{1}{c}{Queue} & \multicolumn{1}{p{0.015\textwidth}}{$15$} &\multicolumn{1}{c}{Queue}& $15$ \\
GEANT & \multicolumn{1}{p{0.01\textwidth}}{$22$} & \multicolumn{1}{p{0.01\textwidth}}{$33$} & \multicolumn{1}{p{0.01\textwidth}}{$10$} & \multicolumn{1}{p{0.01\textwidth}}{$5$} & \multicolumn{1}{c}{Queue} & \multicolumn{1}{p{0.015\textwidth}}{$20$} &\multicolumn{1}{c}{Queue}& $20$ \\
SW & \multicolumn{1}{p{0.01\textwidth}}{$100$} & \multicolumn{1}{p{0.01\textwidth}}{$320$} & \multicolumn{1}{p{0.01\textwidth}}{$30$} & \multicolumn{1}{p{0.01\textwidth}}{$8$} & \multicolumn{1}{c}{(both)} & \multicolumn{1}{p{0.015\textwidth}}{$20$} &\multicolumn{1}{c}{(both)}& $20$ \\

\hline
\textbf{Other Parameters}&\multicolumn{8}{|c|}{
$|\mathcal{T}_a| = 2$, $r_i(a) \in [0.5,1.5]$, $L_{(a,k)} = 10 - 5k$
} \\

\hline
\end{tabular}
\vspace{-0.2\baselineskip}
\caption{Simulated Network Scenarios}
\label{tab_scenario}
\vspace{-0.5\baselineskip}
\end{table}

\begin{figure}[h]
\begin{minipage}{0.49\linewidth}
\includegraphics[width=1\linewidth]{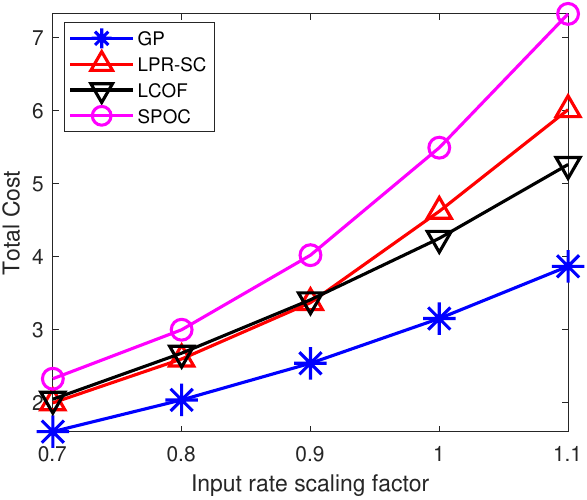}
\caption{$D(\boldsymbol{\phi})$ vs Data input rates}
\label{fig:inputRate}
\end{minipage}
\begin{minipage}{0.49\linewidth}
\includegraphics[width=1\linewidth]{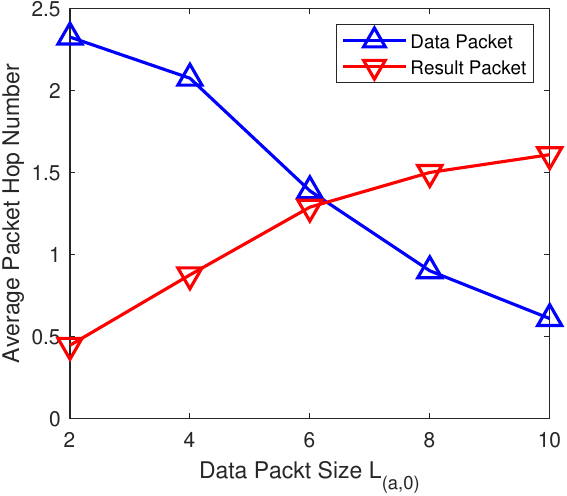}
\caption{Packet hop number vs $L_{(a,0)}$ }
\label{fig:comploc}
\end{minipage}
\end{figure}

We compare \texttt{GP} against: 
\noindent\textbf{SPOC} (Shortest Path Optimal Computation placement) fixes the forwarding variables to the shortest path (measured with marginal cost at $F_{ij} = 0$) and solves the optimal offloading along these paths. 
\noindent\textbf{LCOF} (Local Computation placement Optimal Forwarding) computes all exogenous input flows at their data sources, and optimally routes the result to destinations.
\noindent\textbf{LPR-SC} (Linear Program Rounded for Service Chain) is the joint routing and offloading method by \cite{liu2020distributed} (we extend this method heuristically to service chain applications), which does not consider link congestion. 

Fig.\ref{fig_bar} compares the total cost (normalized according to the worst performing algorithm) across scenarios in Table \ref{tab_scenario}. 
We test both linear cost and convex queueing cost with other parameters fixed in \texttt{SW}, labeled as \texttt{SW-linear} and \texttt{SW-queue}. 
The proposed algorithm \texttt{GP} significantly outperforms other baselines in all simulated scenarios, with as much as $50\%$ improvement over \texttt{LPR-SC} which also jointly optimizes routing and task offloading. 
Case \texttt{SW-linear} and \texttt{SW-queue} suggest that \texttt{GP} promises a more significant delay improvement for networks with queueing effect.
Fig.\ref{fig:inputRate} shows total cost versus exogenous input rates $r_{i}(a)$ in \texttt{Abilene}. The performance advantage of \texttt{GP} quickly grows as the network becomes more congested.
In Fig. \ref{fig:comploc}, we compare average hop numbers travel by data and result packets for \texttt{GP} over the ratio of packet size.
The average computation offloading distance grows when $L_{(a,0)}$ becomes relatively small, i.e., \texttt{GP} tends to offload tasks with larger data size nearer to requester. 

\section{Conclusion} \label{Section:Conclusion}
We propose a joint forwarding and offloading model for service chain applications in CEC.
We formulate a non-convex total cost minimization problem and optimally solve it by providing sufficient optimality conditions. 
We devise a distributed and adaptive online algorithm that reaches the global optimal.
Our method achieves delay-optimal forwarding and offloading for service chain computations, and can be applied to embed computation-intensive complex applications, e.g., DNN, into CEC networks.
Our future work focuses on extending the proposed framework to incorporate general interdependency among tasks, e.g., directed acyclic graphs.

\bibliographystyle{IEEEtran}
\bibliography{References}

\appendix
\subsection{Proof of Theorem \ref{Thm_Sufficient}.}
\label{Proof:Thm_sufficient}
We follow the idea of \cite{gallager1977minimum} and \cite{zhang2022optimal} and generalize to service chain computation applications.
For simplicity, we focus throughout the proof on the situation when $\sum_{j}\phi_{ij}(a,k) = 1$ for every $i$ and $(a,k)$.
The final-stage-destination-node cases with $\sum_{j}\phi_{ij}(a,|\mathcal{T}_a|) = 0$ are omitted since all forwarding fractions must be $0$.

We assume strategy $\boldsymbol{\phi}$ satisfies condition \eqref{Condition_sufficient_chain}. To prove $\boldsymbol{\phi}$ optimally solves \eqref{ServiceChainObj}, let $\boldsymbol{\phi}^\dagger$ be any other feasible strategy such that $\boldsymbol{\phi} \neq \boldsymbol{\phi}^\dagger$, then it is sufficient to show that $T(\boldsymbol{\phi}^\dagger) \geq T(\boldsymbol{\phi})$.
We denote by $f_{ij}^\dagger(a,k)$ and $F_{ij}^\dagger$ the link flows under strategy $\boldsymbol{\phi}^\dagger$, as well as by $g_{i}^\dagger(a,k)$ and $G_i^\dagger$ the computation workloads.

Consider the following flow-based problem, which is equivalent to problem \eqref{ServiceChainObj}:
\begin{align}
    \min_{\boldsymbol{f}} \quad &T_{\boldsymbol{f}}(\boldsymbol{f}) = \sum_{(i,j) \in \mathcal{E}} D_{ij}(F_{ij}) + \sum_{i \in \mathcal{V}} C_i(G_i) \label{proof:flow_domain_problem}
    \\\text{s.t.} \quad & \sum_{j \in \mathcal{V}} f_{ji}(a,k) + g_i(a,k-1) = \sum_{j \in \mathcal{V}} f_{ij}(a,k) + g_i(a,k), \nonumber
    \\ & f_{ij}(a,k) \geq 0, \quad g_i(a,k) \geq 0, \nonumber
\end{align}
where $\boldsymbol{f} = \left[ f_{ij}(a,k),  g_i(a,k)\right]_{(a,k) \in \mathcal{S}, i,j \in \mathcal{V}}$ denotes the flow domain variable.
Note that $T_{\boldsymbol{f}}(\boldsymbol{f})$ is convex in $\boldsymbol{f}$, and the constraints of \eqref{proof:flow_domain_problem} form a convex polytope of $\boldsymbol{f}$. 
Therefore, let $\boldsymbol{f}^\dagger = \left[ f_{ij}^\dagger(a,k),  g_i^\dagger(a,k)\right]$, we know that for any $\alpha \in [0,1]$, $\boldsymbol{f}(\alpha) \equiv (1 - \alpha) \boldsymbol{f} + \alpha \boldsymbol{f}^\dagger$ is also feasible to \eqref{proof:flow_domain_problem}. 
We then let
\begin{equation*}
    \begin{aligned}
        T_\alpha(\alpha) = T_{\boldsymbol{f}}\left(\boldsymbol{f}(\alpha)\right).
    \end{aligned}
\end{equation*}

Then function $T_\alpha(\alpha)$ is also convex in $\alpha \in [0,1]$.
Recall that $T_\alpha(0) = T(\boldsymbol{\phi})$ and $T_\alpha(1) = T(\boldsymbol{\phi}^\dagger)$. To show $T(\boldsymbol{\phi}^\dagger) \geq T(\boldsymbol{\phi})$, it is sufficient to show 
\begin{equation}
    \frac{d T_{\alpha}(\alpha)}{d \alpha} \Big|_{\alpha = 0} \geq 0,
\label{proof:SuffCond_obj}
\end{equation}
where it is easy to see the LHS of \eqref{proof:SuffCond_obj} can be written as
\begin{equation}
\begin{aligned}
    \frac{d T_{\alpha}(\alpha)}{d \alpha} \Big|_{\alpha = 0} = &\sum_{(i,j) \in \mathcal{E}}D_{ij}^\prime(F_{ij})\left(F_{ij}^\dagger - F_{ij}\right) 
    \\ &+ \sum_{i \in \mathcal{V}}C_i^\prime(G_i)\left(G_i^\dagger - G_i\right).
\end{aligned}
\label{proof:SuffCond_derivative}
\end{equation}

For all $i \in \mathcal{V}$, $(a,k) \in \mathcal{S}$, let
\begin{equation*}
    \delta_i(a,k) = \min_{j^\prime \in \left\{0\right\}\cup\mathcal{V}} \delta_{ij}(a,k)
\end{equation*}
denote the minimal augmented marginal among all out-going directions, then by \eqref{Condition_sufficient_chain}, we have
\begin{equation*}
\begin{aligned}
        \frac{\partial T}{\partial t_i(a,k)} &= \sum_{j \in \left\{0\right\}\cup\mathcal{V}} \phi_{ij}(a,k) \delta_{ij}(a,k)
        \\ &= \sum_{j \in \left\{0\right\}\cup\mathcal{V} : \phi_{ij}(a,k) > 0}  \phi_{ij}(a,k)\delta_i(a,k)
        \\ &= \delta_i(a,k)
\end{aligned}
\end{equation*}
and thus for all $i \in \mathcal{V}$, $(a,k) \in \mathcal{S}$ and $j \in \left\{0\right\}\cup\mathcal{V}$,
\begin{equation}
    \delta_{ij}(a,k) \geq \frac{\partial T}{\partial t_i(a,k)}.
\label{proof:SuffCond1}
\end{equation}

Multiply both side of \eqref{proof:SuffCond1} by $\phi_{ij}^\dagger(a,k)$ and sum over $j \in \left\{0\right\}\cup\mathcal{V}$, combining with \eqref{delta}, we have
\begin{equation}
    \begin{aligned}
        \sum_{j \in \mathcal{V}} & L_{(a,k)} D^\prime_{ij}(F_{ij})\phi_{ij}^\dagger(a,k) + w_i(a,k) C^\prime_i(G_i)\phi_{i0}^\dagger(a,k)
        \\ &\geq \frac{\partial T}{\partial t_i(a,k)} - \sum_{j \in \mathcal{V}} \frac{\partial T}{\partial t_j(a,k)}\phi_{ij}^\dagger(a,k) 
        \\ &- \frac{\partial T}{\partial t_i(a,k+1)} \phi_{i0}^\dagger(a,k),
    \end{aligned}
\label{proof:SuffCond2}
\end{equation}

Then multiply both side of \eqref{proof:SuffCond2} by $t_i^\dagger(a,k)$, recall that $f_{ij}(a,k) = t_i(a,k)\phi_{ij}(a,k)$ and $g_i(a,k) = t_i(a,k)\phi_{i0}(a,k)$, we get
\begin{equation}
    \begin{aligned}
        &\sum_{j \in \mathcal{V}} L_{(a,k)}D^\prime_{ij}(F_{ij})f_{ij}^\dagger(a,k) + w_i(a,k)C^\prime_i(G_i)g_{i}^\dagger(a,k)
        \\ &\geq t^\dagger_i(a,k) \frac{\partial T}{\partial t_i(a,k)} - \sum_{j \in \mathcal{V}} t^\dagger_i(a,k) \frac{\partial T}{\partial t_j(a,k)}\phi_{ij}^\dagger(a,k)
        \\ &- {t^\dagger_i(a,k)} \frac{\partial T}{\partial t_i(a,k+1)} \phi_{i0}^\dagger(a,k).
    \end{aligned}
    \label{proof:SuffCond3}
\end{equation}

Recall that for $k = \left|\mathcal{T}_a\right|$ there is no flow forwarded to the computation unit, we write $g_i(a,\left|\mathcal{T}_a\right|) \equiv 0$. 
Sum both sides of \eqref{proof:SuffCond3} over $(a,k) \in \mathcal{S}$, $i \in \mathcal{V}$ and rearrange, it holds that
\begin{equation}
    \begin{aligned}
        &\sum_{(i,j) \in \mathcal{E}} D_{ij}^\prime(F_{ij}) F_{ij}^\dagger + \sum_{i \in \mathcal{V}}C^\prime_i(G_i)G_i^{\dagger}
        \\ &\geq \sum_{i \in \mathcal{V}} \sum_{a \in \mathcal{A}} \sum_{k = 0}^{\left|\mathcal{T}_a\right|} t_i^\dagger(a,k)\frac{\partial T}{\partial t_i(a,k)}
        \\&- \sum_{i \in \mathcal{V}} \sum_{a \in \mathcal{A}} \sum_{k = 0}^{\left|\mathcal{T}_a\right|} {t^\dagger_i(a,k)} \frac{\partial T}{\partial t_i(a,k+1)} \phi_{i0}^\dagger(a,k)
        \\ &- \sum_{a \in \mathcal{A}} \sum_{k = 0}^{\left|\mathcal{T}_a\right|} \sum_{j \in \mathcal{V}} \frac{\partial T}{\partial t_j(a,k)} \left(\sum_{i \in \mathcal{N}(j)} t^\dagger_i(a,k) \phi_{ij}^\dagger(a,k)\right)
    \end{aligned}
    \label{proof:SuffCond4}
\end{equation}

Meanwhile, since the flow conservation
\begin{equation*}
    \sum_{j \in \mathcal{V}} f_{ji}(a,k) + g_i(a,k-1) = \sum_{j \in \mathcal{V}} f_{ij}(a,k) + g_i(a,k)
\end{equation*}
also applies for strategy $\boldsymbol{\phi}^\dagger$, we know for all $j \in \mathcal{V}$, $(a,k) \in \mathcal{S}$,
\begin{equation*}
    \sum_{i \in \mathcal{N}(j)} t_i^\dagger(a,k)\phi_{ij}^\dagger(a,k) = t_j^\dagger(a,k) - g_j^\dagger(a,k-1) - g_j^\dagger(a,k).
\end{equation*}

Substitute above into the right most term of \eqref{proof:SuffCond4}, we get
\begin{equation}
    \begin{aligned}
        &\sum_{(i,j) \in \mathcal{E}} D_{ij}^\prime(F_{ij}) F_{ij}^\dagger + \sum_{i \in \mathcal{V}}C^\prime_i(G_i)G_i^\dagger
        \\ &\geq \sum_{a \in \mathcal{A}} \sum_{k = 0}^{\left|\mathcal{T}_a\right|} \sum_{i \in \mathcal{V}} \frac{\partial T}{\partial t_i(a,k)}g_i^\dagger(a,k-1) 
        \\&+  \sum_{a \in \mathcal{A}} \sum_{k = 0}^{\left|\mathcal{T}_a\right|} \sum_{j \in \mathcal{V}} \frac{\partial T}{\partial t_i(a,k)}g_i^\dagger(a,k)
        \\ & -\sum_{i \in \mathcal{V}} \sum_{a \in \mathcal{A}} \sum_{k = 0}^{\left|\mathcal{T}_a\right|} {t^\dagger_i(a,k)}\frac{\partial T}{\partial t_i(a,k+1)} \phi_{i0}^\dagger(a,k).
    \end{aligned}
\label{proof:SuffCond5}
\end{equation}

We assumed for coherent that $g_i(a,k-1) \equiv r_i(a)$ if $k = 0$, and that $\phi_{i0}(a,\left|\mathcal{T}_a\right|) \equiv 0$, then \eqref{proof:SuffCond5} is equivalent to
\begin{equation}
    \begin{aligned}
        &\sum_{(i,j) \in \mathcal{E}} D_{ij}^\prime(F_{ij}) F_{ij}^\dagger + \sum_{i \in \mathcal{V}}C^\prime_i(G_i) G_i^\dagger
        \\ &\geq \sum_{a \in \mathcal{A}} \sum_{k = 0}^{\left|\mathcal{T}_a\right|} \sum_{i \in \mathcal{V}} \frac{\partial T}{\partial t_i(a,k)}g_i(a,k-1) 
        \\ &+ \sum_{i \in \mathcal{V}} \sum_{a \in \mathcal{A}} \Bigg(\sum_{k = 0}^{\left|\mathcal{T}_a\right|}\frac{\partial T}{\partial t_i(a,k)}g_i^\dagger(a,k-1) 
        \\&- \sum_{k = 0}^{\left|\mathcal{T}_a\right|}\frac{\partial T}{\partial t_i(a,k+1)}g_i^\dagger(a,k)\Bigg)
        \\ &= \sum_{i \in \mathcal{V}} \sum_{(a,k) \in \mathcal{S}} \frac{\partial T}{\partial t_i(a,k)}g_i(a,k-1).
    \end{aligned}
\label{proof:SuffCond6}
\end{equation}

On the other hand, \eqref{proof:SuffCond2} will hold with equality if substitute $\boldsymbol{\phi}^\dagger$ with $\boldsymbol{\phi}$, since \eqref{proof:SuffCond1} hold with equality for $\phi_{ij}(a,k) > 0$. Therefore, the derivation \eqref{proof:SuffCond3}\eqref{proof:SuffCond4}\eqref{proof:SuffCond5} and \eqref{proof:SuffCond6} will hold with equality for $\boldsymbol{\phi}$.
Specifically, substituting $\boldsymbol{\phi}^\dagger$ with $\boldsymbol{\phi}$, we have the following analogue of \eqref{proof:SuffCond6},
\begin{equation}
    \begin{aligned}
        &\sum_{(i,j) \in \mathcal{E}} D_{ij}^\prime(F_{ij}) F_{ij} + \sum_{i \in \mathcal{V}}C^\prime_i(G_i) G_i 
        \\&= \sum_{i \in \mathcal{V}} \sum_{(a,k) \in \mathcal{S}} \frac{\partial T}{\partial t_i(a,k)}g_i(a,k-1).
    \end{aligned}
\label{proof:SuffCond7}
\end{equation}

Subtracting \eqref{proof:SuffCond7} from \eqref{proof:SuffCond6}, we get
\begin{equation*}
     \sum_{(i,j) \in \mathcal{E}}D_{ij}^\prime(F_{ij})\left(F_{ij}^\dagger - F_{ij}\right) + \sum_{i \in \mathcal{V}}C_i^\prime(G_i)\left(G_i^\dagger - G_i\right) \geq 0,
\end{equation*}
which combined with \eqref{proof:SuffCond_derivative} complete the proof of \eqref{proof:SuffCond_obj}.


\end{document}

%% file: macros.tex
\usepackage{cite}
\usepackage{amsmath,amssymb,amsfonts}
\usepackage{graphicx}
\usepackage{textcomp}
\usepackage{xcolor}
\def\BibTeX{{\rm B\kern-.05em{\sc i\kern-.025em b}\kern-.08em
    T\kern-.1667em\lower.7ex\hbox{E}\kern-.125emX}}

\usepackage{amsmath, amssymb}
\usepackage{mathtools}
\usepackage[small]{caption}
\usepackage{amsthm,multirow,color,amsfonts}
\usepackage[ruled,linesnumbered]{algorithm2e}
\usepackage{tabulary}
\usepackage{subfigure}
\usepackage{graphicx}
\usepackage{setspace}
\usepackage{enumerate}
\usepackage{comment}
\usepackage{multicol,lipsum}
\usepackage{cite}
\usepackage{bm}
\usepackage{bbm}
\usepackage[noend]{algpseudocode}
\usepackage{hyperref}
\usepackage{caption}
\usepackage{subcaption}
\usepackage{textcomp}
\usepackage{xcolor}
\makeatletter
\def\BState{\State\hskip-\ALG@thistlm}
\makeatother
	
\newtheorem{theo}{Theorem}

\newtheorem{prop}{Proposition}

\newtheorem{lem}{Lemma}

\usepackage{soul}

\IEEEaftertitletext{\vspace{-1\baselineskip}}